\documentclass[conference,a4paper]{IEEEtran}
\IEEEoverridecommandlockouts
\usepackage{cite}
\usepackage{amsmath,amssymb,amsfonts}
\usepackage{algorithmic}
\usepackage{graphicx}
\usepackage{textcomp}
\usepackage{xcolor}
\usepackage{booktabs}
\usepackage{multirow}
\usepackage{url}
\def\BibTeX{{\rm B\kern-.05em{\sc i\kern-.025em b}\kern-.08em
    T\kern-.1667em\lower.7ex\hbox{E}\kern-.125emX}}
\usepackage{hyperref}
\hypersetup{
    colorlinks   = false,
    citecolor    = black,
    urlcolor    =  black
}
\usepackage{tabularx}
    \newcolumntype{L}{>{\raggedright\arraybackslash}X}
    \newcolumntype{C}{>{\centering\arraybackslash}X}
    \newcolumntype{R}{>{\raggedleft\arraybackslash}X}
\usepackage{subcaption}
\usepackage{tikz}
\usetikzlibrary{arrows.meta, positioning}

\let\oldthebibliography\thebibliography
\let\oldendthebibliography\endthebibliography
\renewenvironment{thebibliography}[1]{%
  \oldthebibliography{#1}%
  \scriptsize
  \setlength{\itemsep}{0pt}%
  \setlength{\parskip}{0pt}%
}{\oldendthebibliography}

\begin{document}

\title{Data-Driven Detection of Volatility Regimes and Sector-Level Asymmetry During Sovereign Default}

\author{
\IEEEauthorblockN{
M.J.S. Ahamed, M.R. Hatheem, M.A.H. Ahamed, U.I. Jayasekara, H.Y. Ahamed, \\
Madara Mendis, Nevidu Jayatilleke, Nisansa de Silva, and Sandareka Wickramanayake
}
\IEEEauthorblockA{\textit{Department of Computer Science and Engineering} \\
\textit{University of Moratuwa}\\
Moratuwa, Sri Lanka \\
\{sabitha.23, hatheemr.23, hamthana.23, upekshaj.23, yusria.23, madara.21, nevidu.25, NisansaDdS, sandarekaw\}@cse.mrt.ac.lk}
}

\maketitle

\begin{abstract}
We present a data-driven framework for detecting changes in stock-market volatility and characterising their effects across industry sectors. The framework identifies volatility regimes directly from historical returns using dynamic programming, validated with a sup-F test and a Markov-switching model. We apply it to 1,559 trading days of the Sri Lankan All Share Price Index (ASPI) and 13 sector indices, covering pre-COVID, COVID-19, sovereign-default, and recovery periods. Within each regime, Student-$t$ EGARCH models quantify volatility response, persistence, and asymmetry between positive and negative shocks. The results reveal a distinct regime-specific pattern: significant negative-shock asymmetry occurs in 9 of 14 series during the sovereign default, compared with none during COVID-19 despite 2.6 times higher volatility than pre-COVID. For the ASPI, volatility persistence decreases from 0.936 during COVID-19 to 0.647 during the default. Cross-sector correlations increase during the default but largely disappear after adjusting for increased variance, providing little evidence of stronger shock propagation. Macroeconomic variables explain less than 4\% of ASPI forecast-error variance at a 30-day horizon. These findings demonstrate how combining automated regime detection with interpretable time-series models can uncover structural changes hidden by aggregate analysis.
\end{abstract}


\section{Introduction}
The macroeconomic crisis that struck Sri Lanka in 2022, a period of hyperinflation, rapid currency devaluation and the depletion of foreign reserves, is a textbook case of sovereign debt collapse in a frontier market. On 12 April 2022 the country suspended payment on its external debt~\cite{cbsl2022}, and the Colombo Stock Exchange (CSE) saw sharp declines and an extended trading halt. Whether the episode also changed how volatility responded to news, rather than simply lowering prices, is the question we address. International precedents suggest such episodes reprice risk broadly: during the Greek sovereign debt crisis, news of a prospective bailout moved European bank equity even for banks with no exposure to Greece, although news about Greece itself did not~\cite{b10}.

While existing research frequently examines how stock indices depreciate during sovereign crises~\cite{b11,b7,b5}, there is a critical gap in analysing how such defaults alter the volatility profile, and specifically the asymmetry, of individual industrial sectors. Traditional portfolio theory suggests that ``defensive'' sectors protect capital during broad market sell-offs. Whether that protection holds during a sovereign default, and whether shocks spread between sectors in ways ordinary correlations would miss, are empirical questions we test directly. Both turn on the behaviour of volatility itself: financial returns cluster, with large movements followed by further large movements~\cite{n1}, producing distinct periods of calm and turbulence that we model explicitly.

We address these questions with a sector-level study of the CSE spanning the default. We first identify volatility regimes from the return series, then fit Exponential Generalized Autoregressive Conditional Heteroskedasticity (EGARCH) models within each regime for the ASPI and 13 sector indices. We test whether cross-sector co-movement rose by more than the increase in variance alone implies, and whether sectors conventionally treated as defensive behaved differently from cyclical ones. Separately, we estimate a Vector Error Correction Model (VECM) linking the index to inflation, interest rates and the exchange rate, and decompose forecast-error variance to quantify how much of the index's movement macroeconomic shocks account for.

\section{Related Work}
The relationship between macroeconomic variables, such as inflation and interest rates, and stock market volatility in emerging economies is heavily documented. Studies of outright sovereign defaults remain scarce, so the closest evidence comes from the European debt crisis. Pereira and Lagoa~\cite{b11} document contagion and flight-to-quality flows among the Portuguese, Greek, Irish and German government bond markets during the European debt crisis, a domino effect of financial stress across neighbouring markets. Extreme inflation and rapid currency devaluation have a substantial impact on emerging equity markets~\cite{b7}. Bollerslev's GARCH model~\cite{n1} captures volatility clustering but treats positive and negative shocks alike. Nelson's EGARCH~\cite{n1,b1} relaxes this by allowing them to have different effects. The difference is called the leverage effect: bad news raises volatility more than good news of the same size. Multivariate approaches such as dynamic conditional correlation~\cite{n4} model co-movement instead, and are not used here.

Evidence on asymmetry in the CSE is mixed, and the disagreement seems to depend on the sample period. Jegajeevan~\cite{jegajeevan2012} finds significant negative asymmetry over 1998--2009 ($\gamma = -0.050$, $p = 0.032$). A later study covering 2018--2022 finds it much weaker, significant only at the 10\% level ($\gamma = -0.038$, $p = 0.063$)~\cite{samarawickrama2023}. Peiris~\cite{peiris2012} finds no leverage effect during the civil war but a significant one afterwards. Taken together, these results suggest that asymmetry in this market depends on which period is studied.

For the Sri Lankan market, Madurapperuma~\cite{b5} uses a VECM, a model that tracks how far a set of variables has drifted from its long-run relationship and how fast it returns, to relate the CSE to domestic interest rates and inflation. The estimated error-correction coefficient is $-0.048$ per month, so about fourteen months pass before half of a deviation is undone.

Existing local research works mainly at the level of the aggregate index, leaving structural shifts in sector-specific volatility during a sovereign collapse largely unquantified. Studies of cyclical and defensive sectors through COVID-19 have focused on behavioural effects such as overconfidence rather than on downside protection~\cite{b14}. The closest prior work is Riyath~\cite{b4}, who fits an ARMA-FIGARCH model to the ASPI and the S\&P SL20 across normal, COVID-19 and crisis periods. That study reports strong long-memory persistence in all three periods, but FIGARCH is a symmetric specification, so it does not test for asymmetry at all, and it covers two indices rather than a sector panel. We ask a different question: how asymmetry differs across sectors within each period.

We address this gap with a sector-level analysis across four data-identified regimes, using the COVID-19 period as a within-sample control for market stress that was not a sovereign default.

\section{Data and Methodology}

\begin{figure}[htb]
\centering
\begin{tikzpicture}[
  node distance = 2.2mm,
  font  = \scriptsize,
  wide/.style = {draw, rounded corners=1.5pt, align=center, inner sep=2pt, text width=79mm},
  core/.style = {draw, thick, rounded corners=1.5pt, align=center, inner sep=2pt,
                 text width=79mm, fill=black!7},
  sub/.style  = {draw, rounded corners=1.5pt, align=center, inner sep=2pt,
                 text width=23mm, minimum height=7mm},
  ar/.style   = {-{Latex[length=1.3mm,width=1.1mm]}}
]
\node[wide] (data)
  {\textbf{Inputs:} CSE ASPI and 13 sector indices (daily) $\cdot$ CBSL CCPI, USD/LKR, SLFR};

\node[wide, below=of data] (prep)
  {\textbf{Preprocessing:} trading-calendar join $\cdot$ CCPI base splice $\cdot$ log transform\\
   liquidity filter (ZRR $\le 5\%$) $\Rightarrow$ 1{,}559 days $\times$ 14 series};

\node[core, below=of prep] (det)
  {\textbf{Regime detection:} ASPI returns};

\node[sub, below=of det] (d2) {sup-F scan\\ simulated null};
\node[sub, left=3mm of d2]  (d1) {Bai--Perron DP\\ + BIC};
\node[sub, right=3mm of d2] (d3) {two-state\\ Markov switching};

\node[wide, below=of d2] (reg)
  {\textbf{Four regimes:} pre-COVID $\cdot$ COVID $\cdot$ default $\cdot$ recovery};

\node[sub, below=of reg] (m2) {Forbes--Rigobon\\ + Fisher $z$\\ $\rightarrow$ contagion};
\node[sub, left=3mm of m2]  (m1) {Student-$t$\\ EGARCH(1,1)\\ $\rightarrow \alpha, \beta, \gamma$};
\node[sub, right=3mm of m2] (m3) {Johansen VECM\\ + FEVD\\ $\rightarrow$ macro share};

\draw[ar] (data) -- (prep);
\draw[ar] (prep) -- (det);
\draw[ar] (det.south) -- (d1.north);
\draw[ar] (det.south) -- (d2.north);
\draw[ar] (det.south) -- (d3.north);
\draw[ar] (d1.south) -- (reg.north);
\draw[ar] (d2.south) -- (reg.north);
\draw[ar] (d3.south) -- (reg.north);
\draw[ar] (reg.south) -- (m1.north);
\draw[ar] (reg.south) -- (m2.north);
\draw[ar] (reg.south) -- (m3.north);
\end{tikzpicture}
\caption{Analysis pipeline. Regime boundaries come from the ASPI returns alone.}
\label{fig:pipeline}
\vspace{-4mm}
\end{figure}
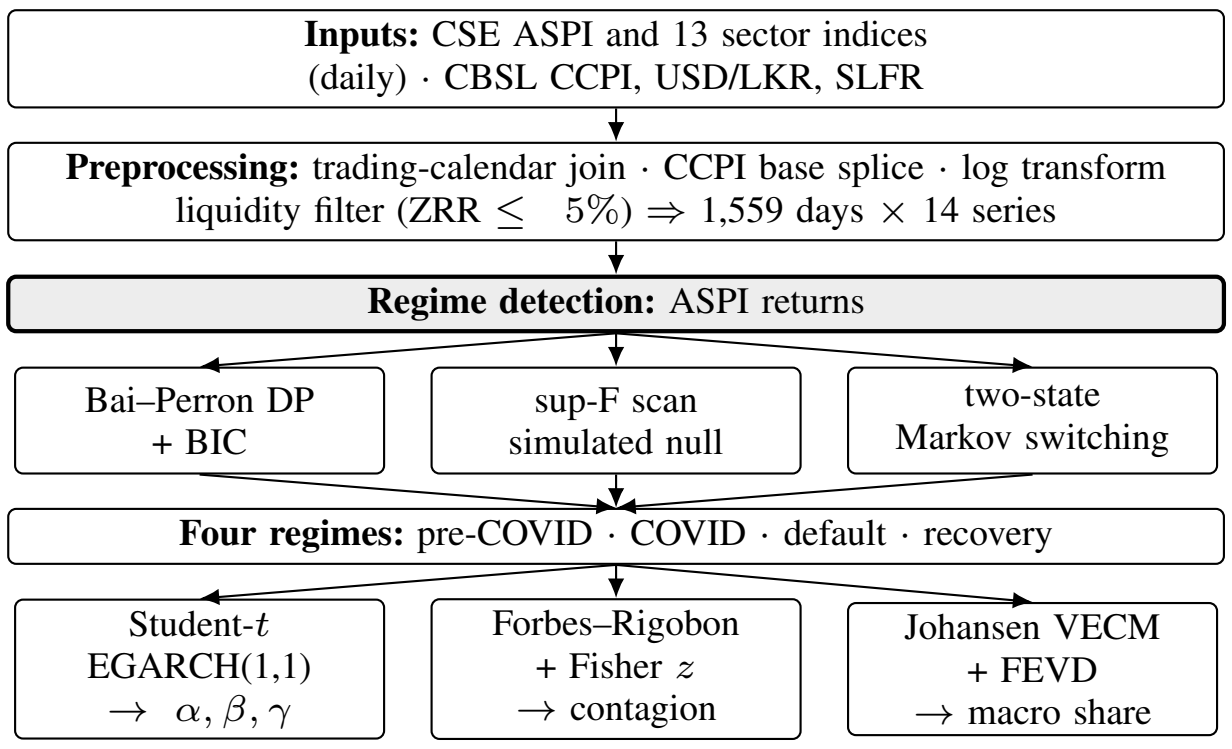

Fig.~\ref{fig:pipeline} summarises the pipeline. Regime boundaries are estimated from the return series alone (Section~III-B) and then held fixed, so every model that follows is fitted within a data-selected regime rather than a policy-calendar one.

\subsection{Data Sourcing and Regime Segmentation}
The dataset comprises 1,559 trading days from 1 January 2019 to 31 August 2025, the end date set by the coverage of the macroeconomic series at the time of data collection. The liquidity filter below uses the full 1,640-day raw sample; the modelling sample is the 1,559 days on which all variables are jointly observed. Regime sample sizes below count returns and therefore sum to 1,558, one fewer than the number of price observations. Daily ASPI and sector index data come from the Colombo Stock Exchange. Concurrent macroeconomic series including the Colombo Consumers' Price Index (CCPI), USD/LKR exchange rate (rupees per dollar), and the Standing Lending Facility Rate (SLFR) were extracted from the Central Bank of Sri Lanka (CBSL)\footnote{Index data: \url{https://www.cse.lk/listed-entities/indices}. Macroeconomic data: \url{https://www.cbsl.lk/eResearch/}}. All estimation is in Python 3.13, using \texttt{statsmodels}~\cite{seabold2010statsmodels} for the VECM and diagnostics, \texttt{arch} for the EGARCH models and \texttt{ruptures} for the break search.\footnote{Pipeline, estimation scripts and filtered datasets: \url{https://anonymous.4open.science/r/CSE-Research-C03B}} Regimes are estimated from the return series itself, by the procedures set out in Section~III-B, which place breaks on 11 February 2020, 25 January 2022 and 6 December 2022. This yields four periods. Each corresponds to an identifiable episode, although no policy date was used to locate them:

\begin{itemize}
    \item \textbf{Pre-COVID (Jan 2019 -- Feb 2020), $N = 265$:} The calmest period in the sample, with ASPI return volatility of 0.56\% per day.
    \item \textbf{COVID-19 (Feb 2020 -- Jan 2022), $N = 438$:} Volatility rises to 1.42\%. The period includes an extended exchange closure in 2020 and the subsequent recovery in nominal prices.
    \item \textbf{Sovereign default (Jan 2022 -- Dec 2022), $N = 200$:} The most volatile period, at 2.19\%. It contains the abandonment of the LKR 200/US\$ peg on 7 March 2022, a 700 basis point policy rate rise in April 2022, and the suspension of external debt servicing on 12 April 2022~\cite{cbsl2022}.
    \item \textbf{Recovery (Dec 2022 -- Aug 2025), $N = 655$:} Volatility falls to 0.92\%, following exchange-rate stabilisation and the approval of a 48-month IMF Extended Fund Facility on 20 March 2023~\cite{imf2023}.
\end{itemize}

\subsection{Change-Point Detection and Validation}
The regime boundaries above are estimated from the ASPI return series by the first of the three procedures below; the other two test whether that segmentation holds up. All three see only the returns and use no information about the policy calendar.

The first searches for multiple breaks in the log variance. Following Bai and Perron~\cite{baiperron1998}, dynamic programming finds the split points minimising the total within-segment sum of squares, with each segment at least 126 trading days long, and the Bayesian Information Criterion (BIC), which penalises each additional break, selects how many. Three breaks minimise the BIC (Table~\ref{tab:breaks}), at 11 February 2020, 25 January 2022 and 6 December 2022.

The second is a sup-F test, which scans every candidate split point and reports the largest F-statistic. It reaches 124.5 at 11 February 2020~\cite{andrews1993}. To judge whether a statistic that large could arise by chance, we simulate the null distribution rather than quote published tables, which have been revised since the original paper: 5{,}000 artificial series of the same length, with constant variance and no break, give 5\% and 1\% cutoffs of 8.62 and 12.19. The observed 124.5 is an order of magnitude above either.

The third is a two-state Markov-switching model~\cite{hamilton1989}, which assigns each day to a high- or low-volatility state. It separates a calm state ($\sigma = 0.64\%$, average run 30 days) from a turbulent one ($\sigma = 2.11\%$, average run 12 days), and assigns 83\% of the estimated default period to the turbulent state, against 40\% of the COVID period and 1\% of the pre-COVID window.

Two features of these boundaries matter. The largest single break is not the default but the onset of COVID-19 in February 2020; treating it as a regime boundary rather than absorbing it into a long pre-crisis window is what makes the COVID period usable as a control. The two default boundaries fall inside the corresponding policy events, the estimated start preceding the abandonment of the currency peg by about six weeks and the estimated end preceding the IMF approval by roughly three months, which suggests markets repriced ahead of the announcements. An independent study places a regime boundary in the same month, though by hand: Riyath~\cite{b4} dates a crisis period from 4 January 2022 on the evidence of fuel and gas shortages, three weeks before the break our search finds.

Table~\ref{tab:breaks} also reports the same estimates under a policy-calendar segmentation, dating the default from the currency float of March 2022 to the eve of the IMF agreement in February 2023. Asymmetry is weaker and less precisely determined there ($\gamma = -0.150$, $p = 0.031$ against $-0.221$, $p = 0.005$), with six of fourteen series significant rather than nine.

\begin{table}[!htb]
\caption{Estimated Variance Breaks and Sensitivity of the EGARCH Results to Segmentation}
\vspace{-4mm}
\begin{center}
\setlength{\tabcolsep}{3pt}
\resizebox{\columnwidth}{!}{%
\begin{tabular}{lcc}
\toprule
\multicolumn{3}{l}{\textbf{Panel A: Break selection by BIC}} \\
\midrule
\textbf{Breaks ($k$)} & \textbf{BIC} & \textbf{Estimated dates} \\
2 & 2710.6 & Feb 2020, Dec 2022 \\
\textbf{3} & \textbf{2693.6} & \textbf{Feb 2020, Jan 2022, Dec 2022} \\
4 & 2696.7 & adds Jan 2021 \\
\midrule
\multicolumn{3}{l}{\textbf{Panel B: ASPI default-regime EGARCH}} \\
\midrule
 & \textbf{Estimated breaks} & \textbf{Policy calendar} \\
\midrule
Default window & 2022-01-25 to & 2022-03-01 to \\
($N$) & 2022-12-05 (200) & 2023-02-28 (234) \\
$\gamma$ ($p$) & $\mathbf{-0.221}$ (0.005) & $-0.150$ (0.031) \\
Sig. series & 9/14 & 6/14 \\
\bottomrule
\end{tabular}%
}\par
\vspace{0.6mm}
{\scriptsize Note: the estimated-break segmentation is used throughout the paper. sup-F for a single variance break is 124.5 at 2020-02-11 (5\% simulated critical value 8.62). Sig. series counts those with a significant negative $\gamma$ at the 5\% level.\par}
\label{tab:breaks}
\end{center}
\vspace{-4mm}
\end{table}

\begin{figure}[!htb]
\centering
\includegraphics[width=\linewidth]{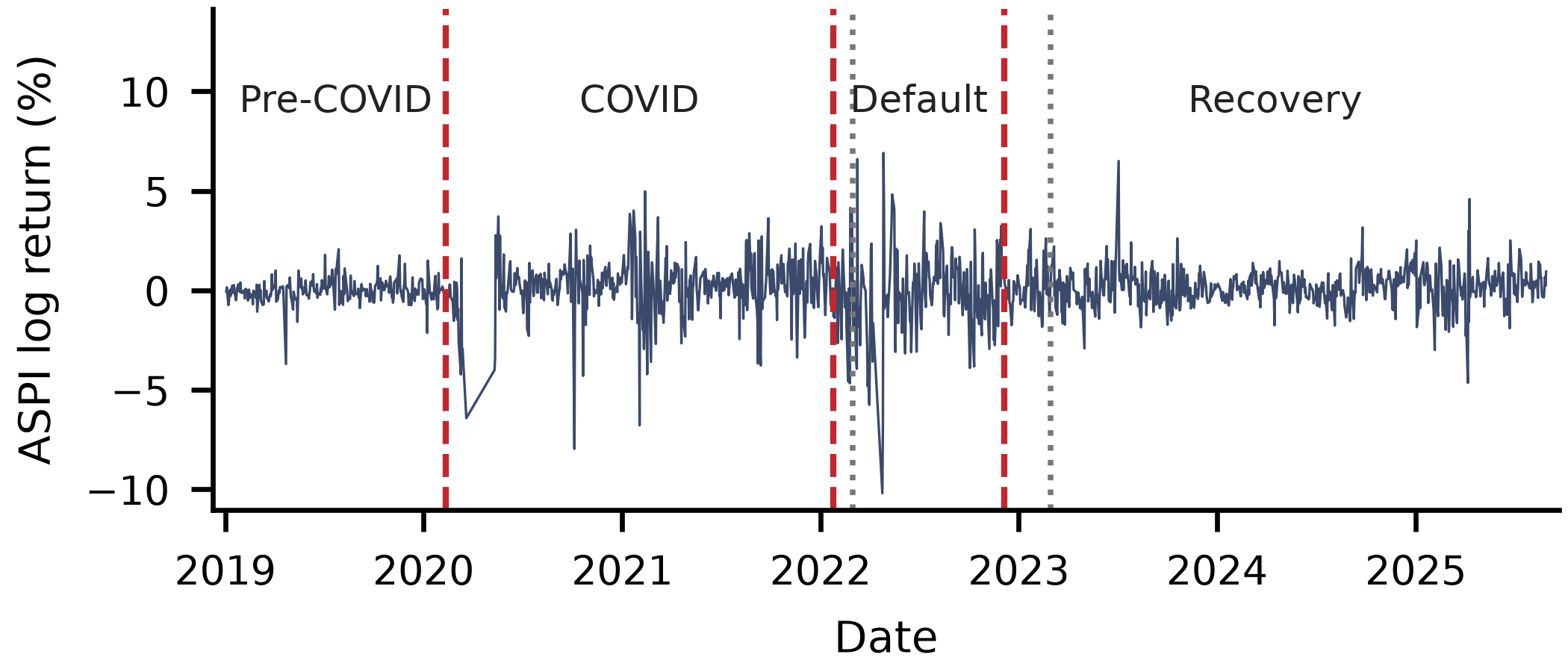}
\caption{ASPI daily log returns. Dashed: estimated variance breaks. Dotted: policy-calendar boundaries.}
\label{fig:breaks}
\vspace{-3mm}
\end{figure}

\subsection{Data Engineering and Frequency Alignment}
Aligning daily market data with monthly macroeconomic series requires three decisions. Daily interest rates and monthly CCPI and USD/LKR values are merged onto the CSE trading calendar by left-join, with intermediate values carried forward. Non-trading days, including the April 2022 market closure, are dropped rather than imputed as zero returns, which would otherwise distort the conditional variance estimates. Finally, the CBSL inflation series changes base from 2013 to 2021 mid-sample; we splice it using a factor derived from the 13-month overlap, giving a continuous series before log transformation.

\subsection{Filtering, Transformation and Stationarity}
Illiquid sectors are screened out by their Zero-Return Ratio (ZRR), the share of days on which the index does not move. Sectors above 5\% are dropped, the worst being Household Products at 20.76\%, because zero-inflated series cause singular-matrix failures in maximum likelihood estimation. Two further sectors, Software \& Services and Utilities, met the threshold but were excluded because they have only 956 trading days of history against 1{,}640 for the rest. This left 13 liquid, representative sectors.

To neutralise hyperinflationary scaling errors, the CCPI, exchange rate and ASPI enter in natural logarithms, so coefficients read as percentage changes.

A series is stationary when its mean and variance do not drift over time, so that it has a level it returns to. Augmented Dickey-Fuller tests establish each series' integration order: how many times it must be differenced, that is replaced by its period-to-period change, before stationarity holds.

The ASPI and all 13 sector log return series are stationary in levels ($p < 0.001$ in every case), so they enter the EGARCH variance equations directly. The index level (ln\_ASPI) and the three macroeconomic series (ln\_CCPI, ln\_Exchange\_Rate and SLFR) are non-stationary in levels but stationary in first differences, and are therefore $I(1)$. This is why we use a VECM rather than regressing the levels on each other directly: two series that both trend upward will appear correlated even when unrelated.

\section{Multivariate Modelling: VECM Results}

Cointegration means that series which individually wander without returning to a fixed level still move together, so that some combination of them is stable. Johansen's trace test~\cite{b2}, on all 1,559 observations with six lags in levels chosen by BIC, finds exactly one such combination: it rejects the hypothesis of none (trace statistic 98.74 against a 5\% critical value of 47.85) but not that of at most one. Rescaled so the index carries a coefficient of one, that combination is $\ln$ASPI $= -9.51\ln(\text{USD/LKR}) + 0.21\,\text{SLFR} + 6.22\ln(\text{CCPI})$. We do not read elasticities off it. The exchange rate and the price level are nearly collinear here (correlation 0.955 in logs), so the fit cannot separate their contributions: the two large coefficients offset one another, and the variance inflation factor on the exchange rate is 20.9 against a conventional threshold of 10. The test establishes that a stable long-run relation exists, not what its separate terms are.

The error-correction term for the ASPI equation is negative and significant at the 5\% level ($\alpha = -0.00125$, $t = -2.06$, $p = 0.039$), implying a half-life of roughly 2.2 years. That estimate should not be read as a property of the market. Running the trace test separately within each regime finds a cointegrating relation during the COVID and default periods but none before COVID or during the recovery, at the 5\% level under either lag rule. We therefore report the full-sample relation as a summary of this period rather than as a stable economic relationship. The residuals show no remaining autocorrelation in any of the four equations (Ljung-Box $p$ between 0.72 and 0.998 at lag 5), so six lags suffice and the variance decomposition that follows rests on an adequately specified model.

Impulse responses to one-standard-deviation shocks show immediate downward pressure on the index from interest-rate and inflation shocks and a lagged, partially corrective path from exchange-rate shocks; the plots are in the accompanying repository.

The index's own past shocks dominate at every horizon we examine. Movements that hit the index and a macroeconomic series on the same day are attributed to the index (a Cholesky ordering, index first), so the macroeconomic shares below are lower bounds. The forecast-error variance decomposition (FEVD) attributes about $1.1\%$ of ASPI forecast-error variance to macroeconomic shocks after five days and $3.9\%$ after thirty, of which the exchange rate contributes $1.85\%$ and inflation $1.70\%$. Around $96\%$ of the variance is therefore internal even at a one-month horizon. Over these horizons the index is driven far more by its own dynamics than by macroeconomic transmission, and we do not read the small macro share as evidence of a strong systemic link at daily frequency.

\section{Cross-Sector Dependence and Contagion Testing}

Log returns are far from normal in every series. The Jarque-Bera test rejects normality for all 13 sectors and for the ASPI ($p < 0.001$), with statistics ranging from 2{,}052 (Consumer Services) to 2.54 million (Real Estate); the latter is driven almost entirely by a single $+46.8\%$ move on 20 March 2023 that is consistent with an index reconstitution. 

Heavy tails like these are what volatility clustering produces: when calm and turbulent stretches alternate, the pooled return distribution has far more extreme days than a normal one allows. That is the case for modelling variance day by day~\cite{b1} rather than treating it as fixed. Volatility clustering is concentrated in the default period, as Fig.~\ref{fig:breaks} shows.

Mean pairwise sector correlation rises from 0.383 before the default to 0.478 during it (Table~\ref{tab:contagion}, Panel A). Read at face value, this is consistent with contagion.

That reading is not available. As Forbes and Rigobon~\cite{n2} show, measured correlations rise automatically when a market becomes more volatile, even with no change in the underlying relationships, so raw comparisons cannot separate contagion from interdependence at higher variance. The problem is acute here: over the same window ASPI variance rises by a factor of 3.44.

We therefore apply their correction, comparing each sector against a market factor built from the equal-weighted average return of the other twelve sectors, so the reference series never contains the sector being tested. Corrected correlations are compared with their pre-default values by Fisher $z$ test (Table~\ref{tab:contagion}).

The result does not support contagion. After correction, one of thirteen sectors shows a significant increase in correlation with the market factor, seven show significant decreases and five are unchanged. This is not an artefact of the market factor: repeating the test against the ASPI gives no significant increases at all, and across all 156 directed sector pairs only eleven (7.1\%) show one. The stronger co-movement seen during the default is therefore explained by the rise in variance, not by a change in how shocks travel between sectors.

Two of the seven decouplers are worth isolating. Healthcare and Telecommunications fall from corrected correlations of 0.459 and 0.498 before the default to 0.221 and 0.259 during it ($p = 0.0008$ and $p = 0.0005$), real decoupling rather than a variance effect. Both are conventionally classified as defensive, and we return to them in Section~\ref{sec:egarch}.
 
\begin{table}[!htb]
\caption{Heteroskedasticity-Adjusted Cross-Sector Correlation Tests}
\vspace{-4mm}
\begin{center}
\resizebox{0.48\textwidth}{!}{
\begin{tabular}{lccc}
\toprule
\multicolumn{4}{l}{\textbf{Panel A: Regime summary}} \\
\midrule
\textbf{Regime} & \textbf{$N$} & \textbf{Mean pairwise $\rho$} & \textbf{ASPI variance} \\
Pre-default & 703 & 0.383 & 1.388 \\
Default & 200 & 0.478 & 4.776 \\
Recovery & 655 & 0.309 & 0.847 \\
\midrule
\multicolumn{4}{l}{\textbf{Panel B: Series with a significant adjusted change}} \\
\midrule
\textbf{Specification} & \textbf{Default} & \textbf{Recovery} & \textbf{Direction} \\
Leave-one-out factor & 1/13 & 6/13 & increase \\
Leave-one-out factor & 7/13 & 3/13 & decrease \\
ASPI as market proxy & 0/13 & 5/13 & increase \\
All directed sector pairs & 11/156 & 40/156 & increase \\
\bottomrule
\end{tabular}
}
\vspace{0.6mm}
{\scriptsize Note: adjusted correlations follow Forbes and Rigobon~\cite{n2}; significance from a Fisher $z$ test at the 5\% level, against the combined pre-default window.\par}
\label{tab:contagion}
\end{center}
\vspace{-4mm}
\end{table}

\begin{table*}[!htb]
\vspace{-2mm}
\caption{EGARCH(1,1) Volatility Dynamics Across Four Estimated Regimes}
\vspace{-4mm}
\begin{center}
\scriptsize
\setlength{\tabcolsep}{3pt}
\begin{tabularx}{\textwidth}{l CCC CCC CCC CCC}
\toprule
\multirow{2}{*}{\textbf{Series}} & \multicolumn{3}{c}{\textbf{Pre-COVID} ($N{=}265$)} & \multicolumn{3}{c}{\textbf{COVID} ($N{=}438$)} & \multicolumn{3}{c}{\textbf{Default} ($N{=}200$)} & \multicolumn{3}{c}{\textbf{Recovery} ($N{=}655$)} \\
\cmidrule(lr){2-4} \cmidrule(lr){5-7} \cmidrule(lr){8-10} \cmidrule(lr){11-13}
& $\alpha$ & $\beta$ & $\gamma$ & $\alpha$ & $\beta$ & $\gamma$ & $\alpha$ & $\beta$ & $\gamma$ & $\alpha$ & $\beta$ & $\gamma$ \\
\midrule
\textbf{ASPI (Systemic)} & 0.15 & 0.918 & 0.028 & 0.48 & 0.936 & 0.006 & 0.48 & 0.647 & \textbf{-0.221}* & 0.40 & 0.848 & -0.078 \\
\textbf{Banks} & 0.48 & 0.816 & 0.117 & 0.42 & 0.943 & -0.024 & 0.55 & 0.797 & -0.150 & 0.30 & 0.948 & -0.022 \\
\textbf{Capital Goods} & 0.13 & 0.887 & 0.075 & 0.35 & 0.942 & -0.019 & 0.59 & 0.685 & \textbf{-0.260}* & 0.33 & 0.710 & -0.044 \\
\textbf{Consumer Durables} & -0.14 & 0.983 & \textbf{-0.041}*$^{\dagger}$ & 0.44 & 0.950 & 0.059 & 0.39 & 0.735 & \textbf{-0.273}* & 0.19 & 0.925 & -0.021 \\
\textbf{Consumer Services} & -0.16 & 0.980 & 0.020$^{\dagger}$ & 0.49 & 0.852 & -0.040 & 0.34 & 0.798 & -0.032 & 0.38 & 0.518 & 0.050 \\
\textbf{Div. Financials} & 0.61 & 0.225 & -0.052 & 0.70 & 0.895 & 0.017 & 0.28 & 0.944 & \textbf{-0.137}* & 0.28 & 0.845 & -0.049 \\
\textbf{Energy} & 0.38 & 0.229 & 0.043 & 0.51 & 0.843 & 0.037 & 0.43 & 0.795 & -0.137 & -0.04 & 0.998 & 0.000$^{\dagger}$ \\
\textbf{Food \& Bev.} & 0.09 & 0.954 & 0.066 & 0.43 & 0.907 & -0.002 & -0.16 & 0.975 & \textbf{-0.119}*$^{\dagger}$ & 0.13 & 0.946 & -0.033 \\
\textbf{Healthcare} & 0.62 & 0.598 & 0.274 & 0.66 & 0.796 & -0.098 & -0.21 & 0.960 & \textbf{-0.250}*$^{\dagger}$ & -0.03 & 0.9996 & 0.019$^{\dagger}$ \\
\textbf{Insurance} & 0.03 & 0.419 & -0.109 & 0.48 & 0.985 & -0.142 & -0.08 & 0.950 & \textbf{-0.420}* & 1.07 & 0.569 & 0.064 \\
\textbf{Materials} & 0.47 & 0.588 & -0.053 & 0.45 & 0.945 & -0.009 & 0.65 & 0.601 & -0.096 & 0.34 & 0.884 & \textbf{-0.071}* \\
\textbf{Real Estate} & 0.07 & 0.977 & 0.000 & 0.39 & 0.893 & -0.080 & 0.29 & 0.818 & \textbf{-0.180}* & 0.20 & 0.698 & 0.018 \\
\textbf{Retailing} & -0.15 & 0.974 & 0.013$^{\dagger}$ & 0.41 & 0.944 & 0.011 & 0.22 & 0.880 & \textbf{-0.202}* & 0.33 & 0.736 & -0.038 \\
\textbf{Telecom} & 0.38 & 0.000 & 0.165 & 0.39 & 0.885 & -0.056 & 0.71 & 0.509 & 0.013 & 0.39 & 0.198 & 0.267 \\
\midrule
\textbf{Series with significant $\gamma<0$} & \multicolumn{3}{c}{\textbf{1 / 14}} & \multicolumn{3}{c}{\textbf{0 / 14}} & \multicolumn{3}{c}{\textbf{9 / 14}} & \multicolumn{3}{c}{\textbf{1 / 14}} \\
\quad excluding $\dagger$ cells & \multicolumn{3}{c}{0 / 11} & \multicolumn{3}{c}{0 / 14} & \multicolumn{3}{c}{7 / 12} & \multicolumn{3}{c}{1 / 12} \\
\bottomrule
\end{tabularx}
\vspace{0.6mm}
{\scriptsize Note: * denotes a statistically significant negative asymmetry coefficient ($p<0.05$). $\dagger$ marks the seven cells whose fitted covariance matrix is numerically singular (standard errors below $10^{-4}$), so their $p$-values are not interpretable. Each cell is fitted from nine starting values, retaining the highest-likelihood result in a plausible parameter region; all 56 converge. Full output is in the accompanying repository.\par}
\label{tab:egarch_full}
\end{center}
\vspace{-6mm}
\end{table*}

\section{Per-Regime Volatility Models: EGARCH}
\label{sec:egarch}

A Student-$t$ EGARCH(1,1) model~\cite{b1} is fitted to the ASPI and each of the 13 sectors within all four estimated regimes, giving 56 fits (Table~\ref{tab:egarch_full}). The specification separates three quantities: $\alpha$, the response of volatility to the size of the latest shock; $\beta$, how long a shock keeps affecting volatility; and $\gamma$, the asymmetry between negative and positive shocks of equal size. A negative $\gamma$ is the leverage effect.

The EGARCH likelihood has multiple local optima, so a single optimisation run on a short sub-sample can settle on a poor solution with implausible parameters. We therefore fit each cell from nine different starting points and keep the best-fitting result that lies in a plausible parameter range. All 56 cells converge under this procedure.

Four estimates deserve comment. The Healthcare and Energy recovery-period $\beta$ values, 0.9996 and 0.9980, sit at the integrated-GARCH (IGARCH) boundary, meaning volatility shocks in those sub-samples decay very slowly. At the other extreme, the Telecommunications pre-COVID $\beta$ is estimated at the lower bound of zero; that cell should be read as no detectable persistence in a 265-day window rather than as a precise estimate. Seven cells, marked $\dagger$ in Table~\ref{tab:egarch_full}, return a numerically singular covariance matrix, with standard errors below $10^{-4}$; their $p$-values cannot be interpreted. All seven also carry a negative $\alpha$, which in EGARCH multiplies $|z_{t-1}| - E|z_{t-1}|$ and should not be read as shock intensity, so we treat both as symptoms of an ill-conditioned fit on a short sub-sample and report the asymmetry counts with and without them. Finally, the Real Estate series contains a $+46.8\%$ move on 20 March 2023 consistent with an index reconstitution rather than a market event; it falls in the recovery regime and inflates that sector's residual variance.

\subsection{What Distinguishes the Default}
Comparing the four regimes separates two effects that a coarser split would merge. Shock magnitude does not distinguish the default. The ASPI $\alpha$ is 0.15 before COVID, 0.48 during COVID and 0.48 during the default: it rises when the pandemic arrives and then simply stays there. Pooling 2019 with the COVID window gives a pre-crisis $\alpha$ of 0.36, which would make the default look like a rise in shock magnitude; separating them shows the rise belongs to COVID.

What does distinguish the default is the collapse in persistence. The ASPI $\beta$ is 0.918 before COVID and 0.936 during it, then falls to 0.647 during the default, a decline significant against the COVID value ($p = 0.018$), and stands at 0.848 in recovery, which is not significantly different from the default level ($p = 0.151$). Volatility during the default was therefore driven by the arrival of new shocks rather than by the lingering memory of old ones, which is characteristic of a market repricing continuously rather than absorbing a single disturbance.

\subsection{Asymmetry Is Specific to the Default}
The clearest result in the panel concerns asymmetry. Significant negative $\gamma$ appears in 1 of 14 series before COVID, 0 of 14 during COVID, 9 of 14 during the default, and 1 of 14 in recovery. Excluding the seven ill-conditioned cells the counts are 0 of 11, 0 of 14, 7 of 12 and 1 of 12: the contrast between the default and every other regime survives, and the single pre-COVID case disappears. For the ASPI itself, $\gamma$ is indistinguishable from zero in the first two regimes ($+0.028$, $p = 0.458$; $+0.006$, $p = 0.890$), significantly negative during the default ($-0.221$, $p = 0.005$), and insignificant again afterwards ($-0.078$, $p = 0.069$).

The COVID regime is what makes this informative. It was a genuine period of market stress: ASPI volatility of 1.42\% per day against 0.56\% before it, a factor of 2.6. Yet not one of the fourteen series developed significant negative asymmetry during it. The default, at 2.19\%, produced nine. Elevated volatility alone therefore does not generate the leverage effect in this market; something particular to the sovereign default does. Our design does not identify the mechanism: rising financial leverage as equity falls, volatility feedback, and loss aversion all produce the same negative $\gamma$. This also explains why estimates in the earlier CSE literature vary in strength: pooling long samples averages one regime in which asymmetry is strong against several in which it is absent.

\subsection{Sector-Class Comparison: Defensive vs Cyclical}
\label{sec:defensive}
Standard portfolio theory says non-cyclical sectors should protect investors during broad sell-offs. That is a claim about a class of sectors, so we test it as one. Sectors are labelled defensive (Healthcare, Food \& Beverage, Telecommunications) or cyclical (the other ten) before inspecting results, and each sector's returns are split into a systematic part, the share of variance a market factor explains, and an idiosyncratic part, the volatility left over (Table~\ref{tab:defensive}). The market factor is the average return of the other twelve sectors. Because the pre-default window pools the calm 2019 period with COVID, we repeat the decomposition against the COVID regime alone; the group comparisons below are unchanged.

Healthcare and Telecommunications combine a fall in explained variance with a rise in idiosyncratic volatility during the default, the pattern that suggests defensive sectors substitute one kind of risk for another. Banks, Capital Goods and Retailing show it too, so it is not distinctively defensive, and Food \& Beverage moves the other way entirely ($R^2$ rising from 0.51 to 0.73).

Comparing the groups directly finds no significant difference on any measure tested (two-sided Mann-Whitney $p \geq 0.371$ for change in explained variance, change in idiosyncratic volatility, mean return and total volatility). With only three defensive sectors surviving liquidity filtering these tests have little power, so this is not evidence that the groups behave alike; we simply cannot separate them with this sample.

What the decomposition does show is narrower, and more favourable to defensive positioning than the raw asymmetry results might suggest. Defensive sectors carried a substantially lower market beta during the default, 0.65 against 1.00 for the cyclical group, and their average return was slightly less negative ($-0.125\%$ per day against $-0.153\%$). The difference in returns is not statistically significant, so we do not claim defensive sectors outperformed. Lower systematic exposure did not mean lower risk: Healthcare and Telecommunications, the two defensive sectors that decoupled, saw residual volatility rise by $75.1\%$ and $61.7\%$. Reduced co-movement with the market did not mean reduced volatility.

\begin{table}[!htb]
\caption{Market-Factor Decomposition by Sector Class, Default vs Pre-Default}
\vspace{-4mm}
\begin{center}
\resizebox{0.48\textwidth}{!}{
\begin{tabular}{lccc}
\toprule
\textbf{Group (default regime)} & \textbf{Market $\beta$} & \textbf{Mean return (\%/day)} & \textbf{Volatility (\%)} \\
\midrule
Defensive ($n=3$) & 0.65 & $-0.125$ & 2.654 \\
Cyclical ($n=10$) & 1.00 & $-0.153$ & 3.107 \\
\midrule
\multicolumn{4}{l}{\textbf{Median by group}} \\
\midrule
\textbf{Measure} & \textbf{Defensive} & \textbf{Cyclical} & \textbf{$p$} \\
Explained variance ($\Delta R^2$) & $-0.053$ & $+0.096$ & 0.371 \\
Idiosyncratic volatility & $+61.7\%$ & $+23.3\%$ & 0.573 \\
Mean daily return & $-0.153$ & $-0.216$ & 0.371 \\
Total volatility (\%) & 2.388 & 2.867 & 0.692 \\
\bottomrule
\end{tabular}
}
\vspace{0.6mm}
{\scriptsize Note: decomposition against a leave-one-out market factor; $p$-values from two-sided Mann-Whitney tests. In the lower block the first two rows are changes from the pre-default window and the last two are levels during the default; the upper block reports group means, the lower block group medians. Per-sector values for all 13 sectors are reported in the accompanying repository.\par}
\label{tab:defensive}
\end{center}
\vspace{-4mm}
\end{table}

\section{Conclusion}
Sri Lanka's 2022 sovereign default changed how volatility responded to news on the Colombo Stock Exchange, not merely its level. Splitting the sample at boundaries estimated from the returns themselves, volatility asymmetry appears in one of fourteen series before COVID, in none at all through the COVID period, in nine of fourteen during the default, and in one of fourteen in recovery. Because the COVID period carried 2.6 times the pre-COVID volatility and produced no asymmetry at all, elevated volatility by itself does not account for the result. Something specific to the sovereign default does, although our design cannot identify the mechanism and a leverage effect is consistent with several.

Separating COVID from the pre-default window also corrects a claim a coarser split would support. Shock magnitude rises when the pandemic arrives and then stays flat through the default, so it is not a signature of the episode. The distinctive change is in persistence, the ASPI $\beta$ falling from 0.936 to 0.647: during the default, volatility was driven by new shocks arriving rather than by old ones failing to decay.

Two things we looked for and did not find are worth stating plainly. Cross-sector correlations do rise during the default, but once corrected for the accompanying rise in variance the increase does not merely disappear, it reverses: one sector shows a significant rise against seven showing significant falls, so the sectors were not transmitting shocks to each other any more than before. And macroeconomic shocks explain under $4\%$ of index variance at a 30-day horizon, so the market stayed largely self-driven even through a sovereign default. On sectors, defensive classification came with a markedly lower market beta, 0.65 against 1.00, but the groups were not statistically distinguishable on any measure tested, and the defensive sectors that decoupled saw idiosyncratic volatility rise sharply: lower co-movement did not mean lower risk.

The study has clear limits. Monthly macroeconomic series are forward-filled to daily frequency, which builds step-like persistence into variables whose daily impulse responses we report. The long-run relation is not stable across regimes, so we report those estimates descriptively. And the sector panel is small, particularly after liquidity filtering. Mixed-frequency estimation would address the first directly. Estimating regime membership jointly with the volatility process, rather than in a prior step as we do, would remove the two-stage design, though neither approach enlarges the sector panel.

{\footnotesize
\bibliographystyle{IEEEtran}
\bibliography{references}
}

\end{document}